# Substrateless metamaterials at mid-infrared frequencies

F. Mattioli[1], M. Ortolani[1], R. Leoni[1], O. Limaj[2] and S. Lupi[2]
[1]CNR-Istituto di Fotonica e Nanotecnologie, via Cineto Romano 42, 00156 Rome (Italy)
[2]CNR-INFM Coherentia and Dipartimento di Fisica, Università di Roma "La Sapienza", Piazzale Aldo Moro 2, I-00185 Roma (Italy)

**Abstract.** We report on the fabrication and mid-infrared transmission properties of free-standing thin metal films, periodically patterned with holes at periods down to 2 microns and area of 3x3 mm$^2$. Square grids were fabricated by electron beam lithography and deep-etching techniques and display substrateless holes, with the metal being supported by a patterned dielectric silicon nitride membrane. The mid-infrared transmission spectra of the substrateless grid display extraordinary transmission peaks and resonant absorption lines with a *Q*-factor up to 22. These spectral features are due to the interaction of the radiation with surface plasmon modes. The high transmittivity and the negative value of the dielectric constant at selected frequencies make our substrateless structures ideal candidates for the fabrication of mid-infrared metamaterials.

The frequency-dependent optical properties of thin metal films patterned at sub-wavelength periods can be exploited to produce optical elements in the infrared (IR) range and beyond [1]. In the wavelength range $\lambda$ = 2-10 μm (mid-IR), which is of high interest for spectroscopy on biomolecules, lithographic techniques are usually employed to obtain structures with micrometric or sub-micrometric periods. The great majority of such structures are obtained by patterning thin metal films evaporated on *i*) IR-transparent polymers or crystals (CaF$_2$, KRS5, etc.) to produce mid-IR polarizers and filters [2] and *ii*) semiconductor substrates (Si, GaAs, InP) for integration of photonic crystal structures with active semiconductor-based devices [3]. Metamaterials (MMs) are a recently introduced class of IR devices based on sub-wavelength periodic structures, showing both high transparency and a negative value of the dielectric constant ε (and/or of the refractive index) [4]. Superfocussing beyond the diffraction limit may be achieved by using MMs [5] so that applications ranging from microspectroscopy to nanolithography and data storage have been envisaged. Recently, mid-IR metamaterials with negative ε at specific frequencies have been fabricated on dielectric substrates and characterized by IR spectroscopy [6-9]. However, for the fabrication of MMs, the presence of a dielectric substrate can influence the phase, the amplitude and the polarization state of electromagnetic waves interacting with the periodic metal structure [10]. Optically active surface plasmon modes, which play a key role in determining the effective optical constants of metamaterials [1,10], are strongly affected by the presence of a substrate on one side of the thin metal film, as the symmetry of the two faces of the film is broken [11,12]. In this paper, we present a technique to fabricate substrateless, large-area (several mm$^2$) thin metal films allowing an entire class of periodic patterns useful to fabricate MMs at mid-IR frequencies. Optical elements in the mid-IR range require periods of the order of 1 μm, precision of the order of 10 nm and patterning of a large area, here provided by electron beam lithography (EBL). We will show how, thanks to substrate removal, negative values of ε can be obtained at specific frequencies with a *Q*-factor up to 22 in a broad mid-IR frequency window with high transmittivity and free of other spectral features.

In order to demonstrate the performances of our substrateless structures at mid-IR frequencies, we selected a design based on a square lattice of holes often used to study the properties of sub-wavelength periodic arrays in the visible range [9-13]. Here, instead, we set the lattice period to *g* = 2.00 μm and we used Fourier-transform spectroscopy (FT-IR) to measure the mid-IR spectrum of resonant surface plasmon modes. The lattice unit was a square-shaped hole of side *g/2*, to produce metal grids. In particular, we demonstrate the existence of resonant surface plasmon absorption at non-normal incidence. Indeed, this resonant absorption were previously observed [8, 14] at far-IR frequencies by free-standing metal grids with periods of hundreds to tens of micrometers and explained in terms of radiation coupling to surface plasmon modes with opposite charge density on the two faces of the metal film. Such resonant absorption features, made allowed by the periodic structure, provide negative values of ε at specific frequencies, which can be used to design MMs [6, 8, 15].

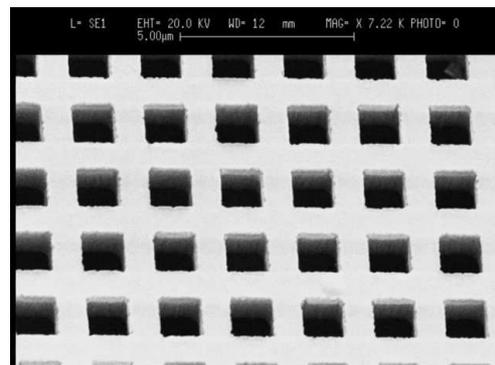

*FIG. 1: Scanning Electron Micrographs of the substrateless grid. The grid covers a 3×3 mm$^2$ area. The sharp corners of the square geometry are due to the EBL technique employed. A Si frame sustains a patterned SiN membrane on top of which an Al layer is deposited.*

We used electron beam lithography and deep etching techniques to fabricate quasi two-dimensional free-standing grids with square-shaped holes of g/2 = 1.00 μm side and 1.00 μm wide metal strips between subsequent holes (Fig. 1). However, we point out that the versatility of EBL allows us to fabricate advanced metamaterial geometries, beyond the square grid, with excellent precision and sharp corners. The metal film is deposited on a self-standing silicon nitride grid, mechanically supported by an open silicon frame. The patterned region of the membrane extends over 3×3 mm$^2$, hence allowing IR transmission spectroscopy with quasi-collimated beams.

The grid fabrication process starts with a <100> silicon wafer covered on both sides with a low-stress, 1 μm thick silicon nitride (SiN) layer. On the front side of the wafer we used EBL followed by dry etching (RIE, Reactive Ion Etching) to obtain the desired pattern on the SiN layer together with alignment markers. A double-side optical mask aligner and a second RIE step are used to obtain clear windows, aligned with the front pattern, on the back SiN layer. A KOH solution (23% in deionized water at 80 °C) is used to etch the Si wafer underneath each pattern and hence obtain a self-standing patterned membrane. 150 nm of Al are then evaporated on the membrane to metallize the surface. As a reference, we fabricated the same grid pattern on a different, double-side polished Si wafer.

The transmittance in the mid-IR were measured in vacuum with FT-IR interferometer (Bruker IFS66v). The radiation was linearly polarized along one side of the squares. The sample was positioned in the focus of an f/4 parabolic mirror and the transmitted beam was collected by a twin f/4 mirror. The collimated beam was then focused into a HgCdTe detector after a 40 cm long path, so that only the radiation transmitted at an angle close to the incoming beam direction was collected, while diffraction spots directed at different angles were not.

Fig. 2 shows the transmittance of two identical grids, one fabricated on a Si substrate and the other fabricated with our substrateless technique. Both spectra display extraordinary transmission peaks related to the excitation of surface plasmons [9-13] and several dips or steps related to power transmission loss into modes diffracted at grazing angle (Wood's anomalies [16]). Apparently, the number, the frequency position and the width of peaks/dips strongly depend on the presence of the substrate. The Wood's anomalies are narrow dips in the transmittance and their frequency position for a square lattice, is given by:

$$\omega(i,j) = \frac{\sqrt{i^2+j^2}}{g\;\varepsilon^{1/2}} \quad (1)$$

where $\varepsilon^{1/2}$ is the index of refraction of the material in which the diffracted beam propagates and $(i, j)$ are integer values defining the two-dimensional diffraction order. In Fig. 2, dashed lines indicate the calculated frequency for the Wood's anomalies related to the first eight diffraction orders in silicon, and the first order in vacuum and SiN: we observe that at least the first four orders diffracted in silicon and the first order diffracted in vacuum perfectly match with the grid-on-Si spectra, as observed before on similar samples [11-13]. Concerning the substrateless grid, there is a broad step corresponding to the (1,0) order in vacuum, but all the other features, including the main dip at 3900 cm$^{-1}$, do not correspond to orders diffracted in vacuum, nor into a SiN substrate ($\varepsilon^{1/2}$ = 2.0). This is not surprising, as the SiN membrane is thin as compared to mid-IR wavelengths and it is patterned by holes, hence it plays no role in the propagation of mid-IR beams. We can therefore conclude that our grid is indeed *substrateless* as far as optical properties are concerned. At odds with the grid-on-Si, the substrateless grid displays a smooth, featureless transmittance in the 3.5 - 10 μm wavelength range, i.e. in the molecular fingerprint region.

The extraordinary transmission peaks are instead to be interpreted as coherent re-emission of excited surface plasmon modes of the metal grid. One may think to the holes as an array of emitting dipoles [10,11]. The re-emission is coherent for all plasmon wavevectors matching the lattice periodicity. Following Ref. 11 we assume a plasmon dispersion relation unperturbed by the holes and we expect peaks at frequencies given by the formula:

$$\varpi(i,j) = \frac{\sqrt{i^2+j^2}}{g}\sqrt{\frac{\varepsilon+\varepsilon_M}{\varepsilon\;\varepsilon_M}} \quad (2)$$

where $\varepsilon_M$ is the (real part of the) dielectric constant of the metal, which is negative and much larger than unity at mid-IR frequencies (much smaller than the plasma frequency of the metal). As a result, the expressions in (1) and (2) are not much different, and the peaks appear at frequencies only slightly lower than the Wood's anomalies. From Eq. 2, we can therefore assign the sharp peaks at 1430 and 2010 cm$^{-1}$ in the spectrum of the grid-on-Si to $\varpi_{Si}(1,0)$ and $\varpi_{Si}(1,1)$ plasmon resonances, and the peaks around 4700-4800 cm$^{-1}$ in both the grid-on-Si and substrateless grid spectra to the $\varpi_{vac}(1,0)$ resonance.

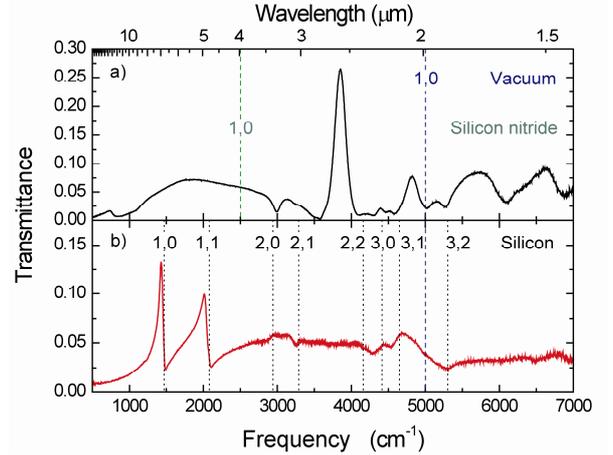

FIG. 2 (color online) Mid-IR transmittance at normal incidence of (a) the substrateless metal grid compared to (b) the same grid fabricated on a silicon substrate. Dashed lines mark the position of Wood's anomalies calculated with Eq. (1) for silicon, vacuum and silicon nitride.

Surprisingly enough, the main peak of the substrateless grid spectrum at $\varpi$ = 3850 cm$^{-1}$ remains out of this assignment, which is rather satisfactory for the grid-on-Si. We can exclude an effect of the SiN membrane, as the nearest frequencies calculated from Eq. 2 are $\varpi_{SiN}(1,1)$ = 3530 cm$^{-1}$ and $\varpi_{SiN}(2,0)$ = 5000 cm$^{-1}$. It seems that while the mid-IR transmittance of the grid-on-Si is well described by the approximated expressions of Eqs. 1 and 2, a deeper analysis is needed for the transmittance of the substrateless grid for $\lambda > g$ ($\omega < 5000$ cm$^{-1}$), where diffraction effects (Wood's anomalies) are not present and resonant surface plasmon modes can be clearly identified [8]. The surface plasmon dispersion for a free-standing square metal grid was first described in Ref. 12. Taking $z$ as the direction orthogonal to the film with $z = 0$ in the mid of the film, the coupling between the two faces of the grid and the structural $\pm z$ symmetry determine the field components of any propagating plasmon. These must be either symmetric ($s$) or antisymmetric ($a$) functions of $z$. The periodic structure couples surface plasmons propagating in the same direction (Bragg reflection) and acts as a perturbation of their dispersion curve effectively realizing a *plasmonic crystal*, with two main consequences for our experiment: i) the degeneracy between $s$- and $a$-modes which exist in a non-patterned film is removed and the $s$- and $a$-modes appear at different frequencies; ii) a gap opens at the plasmonic Brillouin zone edge and two $s$-mode frequencies exist at normal incidence in a given frequency range. This latter fact explains the reason why in the substrateless grid we observe two peaks at 3850 and 4820 cm$^{-1}$ related to the $s_1$- and $s_2$-mode, where the numerical index now runs over the band index of the plasmonic crystal.

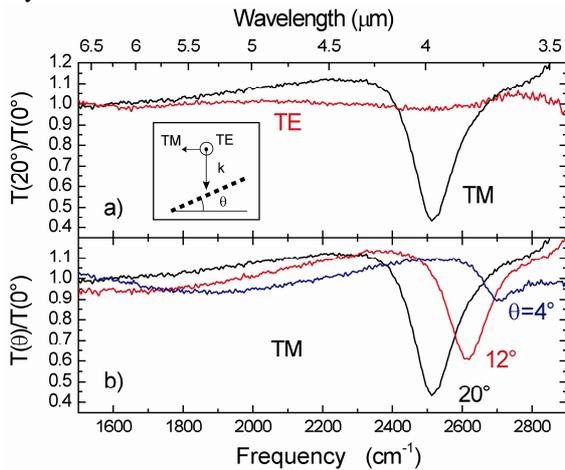

FIG. 3 (color online) (a) Ratio of transmittance at oblique and normal incidence for in-plane (TM) and out-of-plane (TE) polarized radiation in the case of substrateless grid (measurement geometry in the inset). In the TM spectra, the dip close to $\lambda = 2g$ indicates resonant absorption by surface plasmons guided on the grid (leaky mode). (b) Frequency shift as a function of incidence angle due to the leaky mode dispersion

As discussed in the introduction, a key element to obtain a negative value of the dielectric constant is the resonant absorption of energy by surface plasmon modes in the periodic structure of the metamaterial [6, 8, 15]. In our simple square grid geometry, the resonant absorption is provided by $a$-modes that are longitudinal modes and do not couple with radiation impinging at normal incidence. However, for radiation impinging at an angle $\theta$ and polarization parallel to the incidence plane (TM-polarization), the radiation has a field component parallel to $z$ and the dipole moment of $a$-modes of the grid can be directly excited and behaves then as leaky mode. Following Ref. 8, resonant absorption by the lowest-energy $a_1$-mode is expected at a wavelength significantly larger than the period $g$, far from other resonances (see Fig. 2a). This fact is important for the design of MMs, as the observation of $a$-modes can be considerably hindered by Wood's anomalies and/or $s$-modes in substrate-based metamaterials.

We then measured transmittance $T(\theta)$ of the substrateless grid rotated by an angle $\theta$ with respect to the beam direction. More interestingly, an absorption feature, not visible at normal incidence, clearly appears around 2500 cm$^{-1}$. In Fig. 3a we show the transmittance ratio $T(\theta)/T(0°)$ for the substrateless grid of both TM and TE polarizations with $\theta = 20°$, in the 1500-2900 cm$^{-1}$ range. A clear dip appears only in the TM spectrum at a frequency slightly dependent on $\theta$ but close to 2500 cm$^{-1}$ ($\lambda \sim 4$ μm). The $\theta$-dependence of the TM spectra is shown in Fig. 3b for selected values of $\theta$, but the dip was found for any $|\theta| > 0°$. The dip becomes sharper and shifts to lower frequency for increasing $|\theta|$, while an overshoot appears. The $\theta$-dependent feature in the spectra is consistent with the excitation of the $a_1$-mode of the metal grid. Both the dip and the overshoot are predicted in Ref. 14 and are due to a Fano interference between the directly transmitted beam and radiation re-emitted by the excited leaky modes. Within this interpretation, the shift of the resonant frequency with increasing $\theta$ shown is due to the change in the wavevector module $q$ of the $a_1$-mode [14] and to its dispersion relation $\varpi(q)$, whose detailed analysis will be reported in a forthcoming paper. Here, we point out that our substrateless structure provides a resonant absorption with a remarkable $Q$-factor of $\sim 22$ and a certain degree of mechanical tunability, which is an ideal building block for MMs design [4].

In conclusion, we presented a novel fabrication technique of substrateless sub-wavelength metal structures with periods in the few-micron range and we performed a spectroscopic analysis in the mid-infrared. We compared its transmission spectra to those of the same grid fabricated on a silicon substrate and interpreted the main features as excitation of surface plasmon modes. Thanks to substrate removal, resonant absorption was obtained at specific frequencies with a $Q$-factor up to 22 in a broad mid-IR frequency window with high transmittivity and free of other spectral features. The technique here presented can be used to fabricate and test MMs at mid-infrared frequencies.

We thank S. Selci and A. Dilellis of the SERENA /BepiColombo team for discussions. One of us (F.M.) acknowledge the support of the ASI/INAF contract I/090/06/0

[1] W. L. Barnes, A. Dereux and T. W. Ebbesen, Nature **424**, 824 (2003)
[2] J. P. Auton, Appl. Optics **6**, 1023 (1967).
[3] R. Colombelli, K. Srinivasan, M. Troccoli, O. Painter, C. F. Gmachl, D. M. Tennant, A. M. Sergent, D. L. Sivco, A. Y. Cho, and F. Capasso, Science **302**, 1374 (2003).
[4] V. M. Shalaev, Nature Photonics **1**, 41 (2007).
[5] Xiang Zhang and Zhaowei Liu, Nature Materials **7**, 435 (2008).

**[6]** Z. Haoa, M. C. Martin, B. Harteneck, S. Cabrini and E. H. Anderson, Appl Phys. Lett. **91**, 253119 (2007)
**[7]** S. Zhang, W. Fan, B. K. Minhas, A. Frauenglass, K. J. Malloy, and S. R. J. Brueck, Phys. Rev. Lett. **94,** 037402 (2005)
**[8]** T. Ganz, M. Brehm, H. G. von Ribbeck, D. W. van der Weide and F. Keilmann, New Journal of Physics **10,** 123007 (2008).
**[9]** G. Dolling, M. Weg, A. Schädle, S. Burger and S. Linden, Appl. Phys. Lett. **89**, 231118 (2006).
**[10]** F. J. Garcia de Abajo, Rev. Mod. Phys. **79**, 1267 (2007).
**[11]** H. F. Ghaemi, Tineke Thio, D. E. Grupp, T. W. Ebbesen, H. J. Lezec, Phys. Rev. B, **58**, 6779 (1998).
**[12]** D. E. Grupp, H. J. Lezec, T. W. Ebbesen, K. M. Pellerin and Tineke Thio, Appl. Phys. Lett. **77**, 1569 (2000).
**[13]** W. L. Barnes, W. A. Murray, J. Dintinger, E. Devaux, and T. W. Ebbesen, Phys. Rev. Lett. **92,** 107401 (2004)
**[14]** R. Ulrich, "Modes of propagation on an open periodic waveguide for the far-infrared", from Optical and Acoustical Microelectronics, ed. By Jerome Fox, Microwave Research Inst. Symp. Ser. Vol. 23, Polytechnic Press, New York (1974).
**[15]** S. Zhang, W. Fan, N. C. Panoiu, K. J. Malloy, R. M. Osgood, and S. R. J. Brueck, Phys. Rev. Lett. **95**, 137404 (2005)
**[16]** R. W. Wood, Philos. Mag. 4, 396 (1902); R. W. Wood, Phys. Rev. 48, 928 (1935).